The manuscript on the following 2 pages is designated for the

7th Triennial Special Issue of the IEEE Transactions on Plasma Science

"Images in Plasma Science"

scheduled for publication in August 2014

The manuscript length is limited to 2 pages only.  It was submitted for peer review on 2013-11-21.  Receipt was acknowledged on 2013-11-21 and the manuscript number **TPS7217** was assigned.

# Unusual Cathode Erosion Patterns Observed for Steered Arc Sources

Jonathan Kolbeck and André Anders

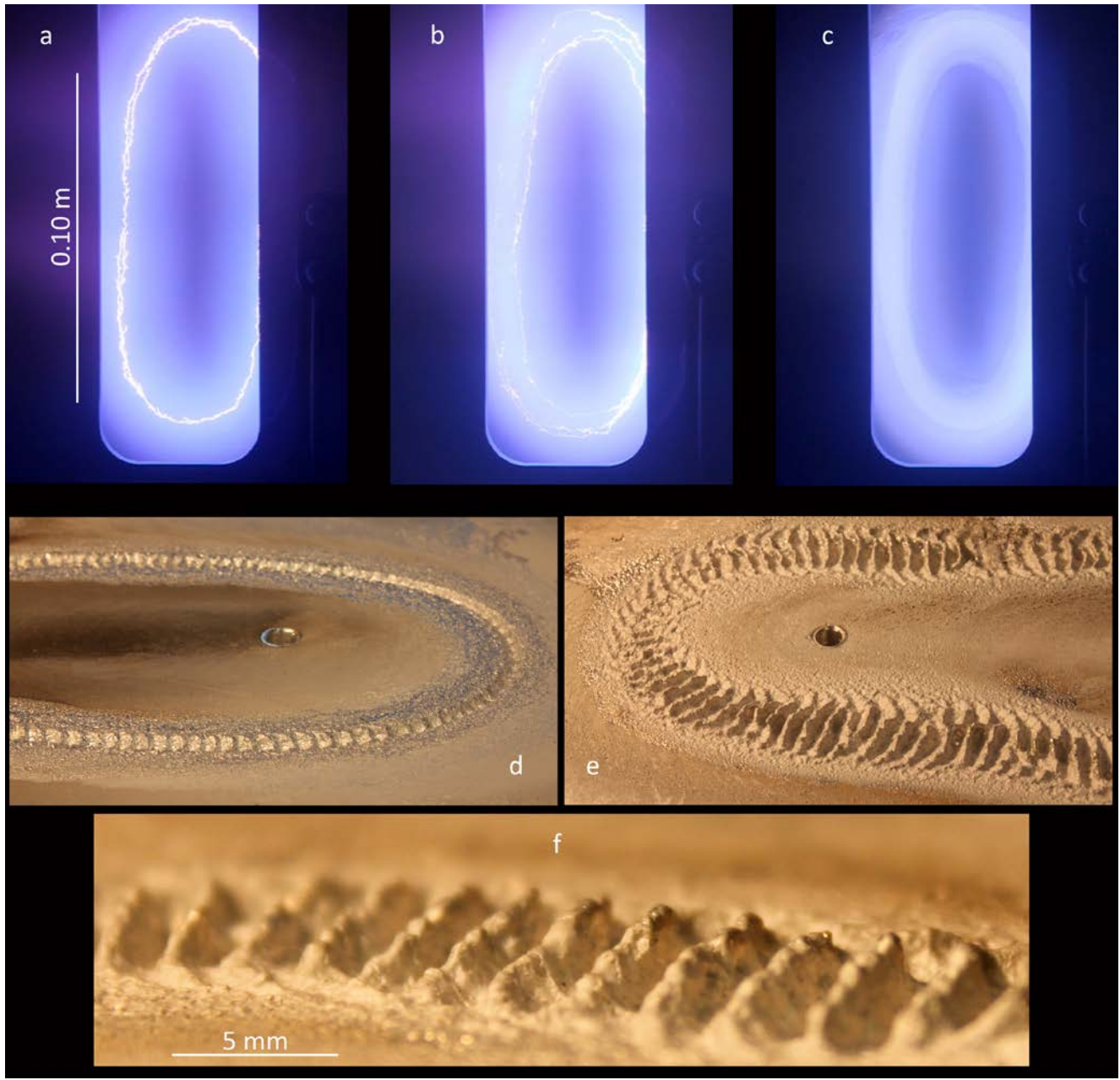

Fig. 1. (a) View onto part of the aluminum cathode through an elongated opening in the anode of a steered arc source operating with a current of 70 A; one can discern the "racetrack" on which arc spots burn, here in 0.5 Pa of pure argon; (b) same but this image was taken about 1 second after oxygen was added to the gas; (c) same as (a) but 10 seconds after oxygen was added; (d) erosion pattern of a cathode operating in pure argon; (e) erosion pattern of a cathode operating in argon-oxygen gas mixture; (f) enlarged detail of (e).

Manuscript received …. November 2013.
K. Kolbeck and A. Anders are with the Lawrence Berkeley National Laboratory, 1 Cyclotron Road, MS 53, Berkeley, California 94720, USA;
Work at LBNL was supported by U.S. Department of Energy under Contract No. DE-AC02-05CH11231.
Publisher Identifier S XXXX-XXXXXXX-X

*Abstract* – A cathodic arc source with a magnetron-type magnetic field was investigated for stability, erosion, and compatibility with a linear macroparticle filter. Here we report about unusual arc spot erosion patterns, which were narrow (~ 2 mm) with periodic pits when operating in argon, and broad (~ 10 mm) with periodic groves and ridges when operating in an argon and oxygen mixtures. These observations can be correlated with the ignition probability for type 2 and type 1 arc spots, respectively.

The study of erosion of cathodes by arc spots has a remarkably long history, which can be traced back to Priestley's work [1, 2]. Two main types of arc spots have been identified for vacuum arcs: small, faint, mobile type 1 spots on contaminated or oxidized metal surfaces, and larger, brighter, slower type 2 spots on clean metal surfaces [3]. Type 1 also occurs when the arc operates in a "reactive" process gas, while type 2 appears not only in vacuum but also in the presence of noble gases.

Type 1 and 2 spots can be best comprehended in terms of arc spot ignition probability [4], which is governed by a strong electric field (~$10^9$ V/m) on the surface. The field leads to field-emission of electrons, which in turn causes localized heating and non-linear amplification of the emission process. Dense plasma of the cathode material is produced, generally containing multiply charged ions. The dense plasma causes contraction of the sheath thickness between plasma and the cathode, thereby enhancing the surface electric field and promoting conditions for ignition of new plasma-producing cathode spots. Field enhancement by non-metallic surface features promotes the ignition of arc spots of type 1 at relatively large distances (e.g. ~100 μm) from existing spots. New and old spots are electrically in parallel, and the older, having higher impedance due to more (insulating!) vapor, is overtaken by the newer. This appears as spot motion. Type 2 spots in contrast, on clean surfaces, rely on the nearby plasma of existing spots to ignite new emission centers. The step width of type 1 spots is therefore much greater than the step width for type 2 spots. The apparent spot velocity for type 1 spots is higher, and their lifetime shorter. All this is well documented in the literature, see overviews [4, 5] and references therein.

A major drawback of vacuum or cathodic arcs is the production of cathode droplets, a.k.a. macroparticles. This feature is important when considering the erosion patterns observed in our cathodic arc source. This arc source has a magnetic field of the "arch type" which is essentially the same structure used in sputtering magnetrons. In fact, this arc source could also be used for magnetron sputtering provided an appropriate magnetron power supply was connected. The magnetic field had its south pole in the center, with the north pole around the periphery. The electrons' $\mathbf{E}\times\mathbf{B}$ drift was thus counterclockwise, and the apparent spot motion was clockwise, known as the "retrograde" motion [6].

Figure 1 shows several aspects of our observations. All images were taken with a Canon Rebel XTi single-lens reflex camera featuring a 4752x3168 pixel CMOS sensor. Images (a)-(c) were extracted from a video, taken with the same camera, each frame having 640x480 pixel at a frame rate of 30 fps, i.e. the exposure time was about 30 ms. In image (a) one can see bright arc traces: the spot moves approximately 2 times the length of the 0.3 m long "racetrack", giving an apparent velocity of 15-20 m/s. When adding the oxygen flow, at an Ar:$O_2$ flow rate of about 2:1, the arc spot appearance changes within 1 second, image (b), to becomes practically diffuse (c), although one can still see tiny spots. In agreement with literature ([4] and references therein), we have many and fast moving arc spots. Images taken with the highest shutter speed of 1/4000 seconds suggest an apparent velocity of about 100 m/s.

The erosion patterns are striking. First, in both cases we see periodic patterns indicative that the spot ignition probability for magnetically steered arcs is not evenly distributed - not only due to the magnetic field but also due to feedback from surface features. In the case of pure argon, racetrack erosion is along a narrow path (d), and distinctly wider when oxygen is added (e). Most remarkably, cathode material is redistributed, not just eroded, as evident by the build-up of ridges, image (f), between the most eroded zones. Very likely, deposition of macroparticles contributes to this re-distribution. The height of the re-distributed material can exceed 1 mm over the original surface, i.e. remarkably 2-3 orders of magnitude greater than the size of the spot erosion craters.